\documentclass[aps,prl,twocolumn,superscriptaddress]{revtex4}
\usepackage{graphicx}
\begin{document}

% Use the \preprint command to place your local institutional report
% number in the upper righthand corner of the title page in preprint mode.
% Multiple \preprint commands are allowed.
% Use the 'preprintnumbers' class option to override journal defaults
% to display numbers if necessary
%\preprint{}

%Title of paper
\title{Excitation dependence of resonance line self-broadening at different
atomic densities}

\author{Hebin Li}

\affiliation{
    Department of Physics and Institute for Quantum Studies,
    Texas A\&M University,
    College Station, Texas 77843-4242
}

\author{Vladimir A. Sautenkov}
\affiliation{
    Department of Physics and Institute for Quantum Studies,
    Texas A\&M University,
    College Station, Texas 77843-4242
}

\affiliation{
   P.N. Lebedev Institute of Physics, Moscow 119991, Russia
}

\author{Yuri V. Rostovtsev}
\affiliation{
    Department of Physics and Institute for Quantum Studies,
    Texas A\&M University,
    College Station, Texas 77843-4242
}

\author{Marlan\ O.\ Scully}
\affiliation{
    Department of Physics and Institute for Quantum Studies,
    Texas A\&M University,
    College Station, Texas 77843-4242
}

\affiliation{ Princeton Inst. for the Science and Technology of
Materials and Dept. of Mech. \& Aerospace Eng., Princeton
University, 08544 }

%Collaboration name if desired (requires use of superscriptaddress
%option in \documentclass). \noaffiliation is required (may also be
%used with the \author command).
%\collaboration can be followed by \email, \homepage, \thanks as well.
%\collaboration{}
%\noaffiliation

\date{\today}

\begin{abstract}
We study the dipole-dipole spectral broadening of a resonance line
at high atomic densities when the self-broadening dominates. The
selective reflection spectrum of a weak probe beam from the
interface of the cell window and rubidium vapor are recorded in the
presence of a far-detuned pump beam. The excitation due to the pump
reduces the self-broadening. We found that the self-broadening
reduction dependence on the pump power is atomic density
independent. These results provide experimental evidence for the
disordered exciton based theory of self-broadening, and can be
useful for the description of the interaction of a strong optical
field with a dense resonance medium.
\end{abstract}

% insert suggested PACS numbers in braces on next line
% insert suggested keywords - APS authors don't need to do this
%\keywords{}

%\maketitle must follow title, authors, abstract, \pacs, and \keywords
\maketitle

% body of paper here - Use proper section commands
% References should be done using the \cite, \ref, and \label commands

\section{Introduction}

\ For many applications and fundamental physics it is necessary to
know the non-linear optical response of a resonance atomic gas under
conditions when dipole-dipole interactions between atoms in the
ground and excited states can not be neglected.

\ There was a common opinion that calculations of self-broadening
can be performed by using the two-particle approximation in the fast
collision limit (impact collisions) as well as in the opposite
limit, static interactions
\cite{Lewis,Vdovin,movre1980,Mukamel1994}. In Ref.
\cite{Mukamel1994}, the theory of self-broadening is developed on
the basis of disordered exciton in a dense resonance medium, in
which many particle interaction should be taken into account. By
using this model it was shown that the self-broadening is a
combination of collision and static atomic interactions. A ratio of
the static width to the collision width is independent on atomic
density in a wide range where thermal motion of atoms can be
neglected.

\ It has been shown that the probing of the homogeneous and
inhomogeneous contributions to the linewidth can be performed
efficiently by nonlinear optical methods such as photon echoes and
hole burning \cite{Mukamel1991,Mukamel1995}. The inhomogeneous
component of the spectral line could be sensitive to the optical
saturation. Recently by using time resolved femto-second
spectroscopy, the non-Markovian collision dynamics and the
bi-exponentional correlation of energy level fluctuations has been
observed in a dense potassium vapor and simulations of molecular
dynamic are in good agreement with experimental results
\cite{Cundiff2008}. The slow exponential component is attributed to
long-range resonant attraction in a dense atomic vapor. By using CW
pump-probe technique, the excitation dependence of the
self-broadening is observed in rubidium \cite{VAS1996, HebinLi2008}
and potassium \cite{VAS1999,VAS2008} vapors. Nevertheless, note that
in these papers the measurements have been performed only at
selected atomic densities.

%\ In the current paper we have studied excitation dependence of
%self-broadening of a resonance atomic line at different atomic
%densities in the range where self-broadening of atomic line is
%stronger than Doppler broadening.

\ In the current paper the selective reflection spectrum from the
interface between the cell window and rubidium vapor are recorded in
the presence of a far-detuned pump beam. We have studied the
excitation dependence of self-broadening of a resonance atomic line
at different atomic densities in the range where self-broadening of
atomic line is stronger than Doppler broadening. The excitation
dependence of self-broadening is found to be independent on the
atomic density. These results support the disorder exciton based
theory of self-broadening\cite{Mukamel1994}.

\section{Experiment}
% Put \label in argument of \section for cross-referencing
%\section{\label{}}

\begin{figure}[htb]
\includegraphics[width= 1\columnwidth]{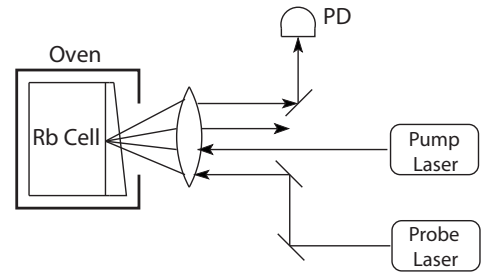}
\caption{The experimental schematic. Pump and probe laser beams are
focused on the interface between rubidium vapor and window. The
reflection is recorded by a photodetector.} \label{fig:setup}
\end{figure}

\begin{figure}[htb]
\includegraphics[width= 0.7\columnwidth]{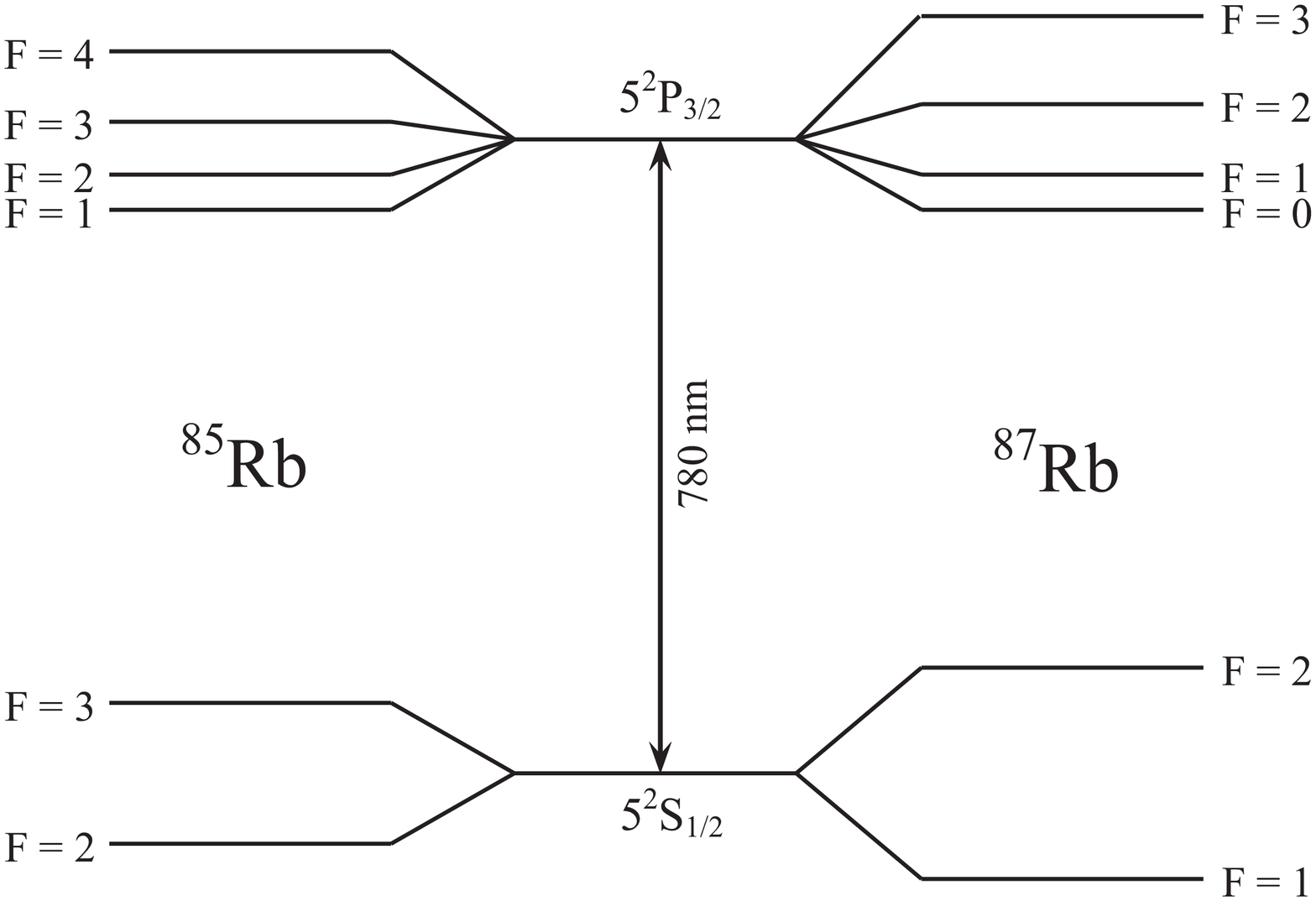}
\caption{The energy diagram of the $D_2$ line of $^{85}$Rb and
$^{87}$Rb. The hyperfine splittings of the excited states are less
than the Doppler width (0.5 GHz) at room temperature and are not
resolved in our experiment.} \label{fig:levels}
\end{figure}

The experiment was performed with a pump-probe scheme shown in Fig.
\ref{fig:setup}. The reflectivity and frequency-modulated (FM)
reflectivity spectra of a dielectric-vapor (rubidium) interface were
measured at the $5^2S_{1/2}\rightarrow 5^2P_{3/2}$ transition ($D_2$
line, shown in Fig. \ref{fig:levels}) of rubidium atoms. Please note
that the excited state hyperfine structure is not resolved in our
experiment since the splittings are less than the Doppler width (0.5
GHz) at room temperature. The probe laser is a free running diode
laser (linewidth 20 MHz) which can be scanned over 30 GHz around
$D_2$ transition. The frequency of the probe laser is calibrated by
reference to the absorption of a rubidium cell at room temperature.
The power of the probe beam is small enough (less than 100 $\mu$W)
such that no saturation effects need to be considered. The pump
laser (an extended cavity diode laser) provides a laser beam with
power up to 180 mW, and it is far-tuned (20 GHz) to the red wing of
$D_2$ line to avoid coherent effects. The frequency of the pump
laser is determined by beating with the probe laser. Both pump and
probe beams are focused down to a spot with 100 $\mu$m diameter by a
15 cm lens, and they are overlapped at the inner surface of the cell
window.

The cell contains natural abundance of rubidium vapor. The cell is
made of sapphire, and the windows are Garnet crystal which is free
of birefringence. The cell can be heated to reach a high atomic
density (N$\approx 10^{17}$ cm$^{-3}$). The windows are slightly
wedged in order to separate the reflections from two surfaces.

\begin{figure}[htb]
\includegraphics[width= \columnwidth]{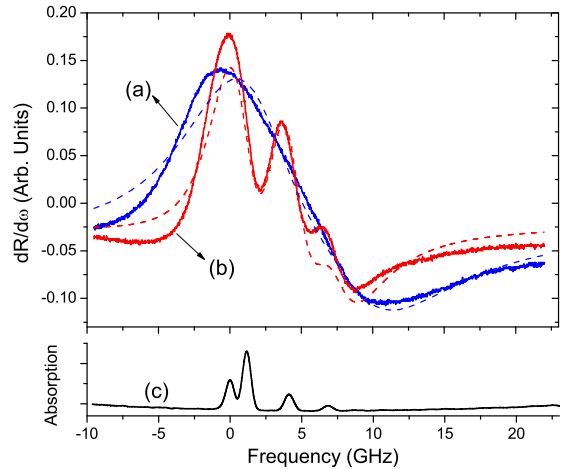}%
\caption{(Color online) FM reflectivity spectra at atomic density
N=$1.3\times 10^{17}$ cm$^{-3}$. Blue (a) and red (b) solid curves
correspond to the cases without pump laser and with pump laser
(P=180 mW), respectively. The dashed curves are corresponding
fitting results. Curve (c) is the absorption spectra of a reference
Rb cell, where the ground state hyperfine splitting is resolved.}
\label{fig:narrow}
\end{figure}

The reflected beam from the interface between the rubidium vapor and
the window is sent to a photodetector (PD). The signal from the
photodetector is processed by a lock-in amplifier while we frequency
modulate the probe laser with modulation depth of 37 MHz at
frequency of 8 kHz. FM reflectivity spectra is used to improve the
signal to noise ratio in our experiment, and it can reveal subtle
details of change in reflectivity. A typical FM reflectivity
spectrum is shown in Fig. \ref{fig:narrow} as curve (a) which was
obtained at the atomic density N=$1.3\times 10^{17}$ cm$^{-3}$. The
dipole-dipole interaction and collision broadening dominate at this
atomic density, the spectral width due to self-broadening is larger
than the ground state hyperfine splitting and the ground state
hyperfine structures in reflectivity spectra that can be seen at low
atomic density are not resolved. While we apply the pump laser beam,
atoms are partially excited and the dipole-dipole interaction is
reduced. Thus, the self-broadened line width is also reduced
\cite{VAS1996,HebinLi2008,VAS1999,VAS2008}, and the ground state
hyperfine structures start to be revealed as described in
\cite{VAS1999,HebinLi2008}. The narrowed FM spectrum with pump power
P=180 mW is shown as curve (b) in Fig. \ref{fig:narrow}.

As described in \cite{VAS1999}, the reflection spectra as well as
the FM reflection spectra can be interpreted in terms of the
dielectric coefficient of atomic vapor. Taking into account the
excitation, the dielectric coefficient $\epsilon$ of a two-level
atomic system is given as

\begin{equation}
\epsilon (\omega)=1+\frac{k\eta N}{\Delta \omega+\Delta \Omega - i
\Gamma},
\end{equation}
where $N$ is the atomic density, $\Delta \omega$ is the frequency
detuning, $\Delta \Omega$ is the overall line shift which includes
Lorentz and non-Lorentz shift
\cite{Friedberg1973,Manassah1983,Maki1992}, and $\Gamma$ is the
self-broadened linewidth. The constant $k$ is given by
$k=fcr_e\lambda$, where $f$ the oscillator strength of transition,
$r_e$ the classical radius of electron, $\lambda$ the wavelength of
transition, and $c$ is the speed of light in vacuum. An excitation
factor $\eta$ is defined as the fractional population difference
between ground and excited states
\begin{equation}
\eta=\frac{N_g-N_e g_g/g_e}{N},
\end{equation}
where $N_g$ and $N_e$ are the ground and excited state atomic
densities respectively, $g_g$ and $g_e$ are the degeneracies of the
ground and excited states respectively. Maximum excitation
corresponds to $\eta=0$ and zero excitation $\eta=1$. Using this
expression for the dielectric coefficient $\epsilon$ and Fresnel
formula, we are able to calculate the reflectivity and FM spectra
which is the derivative of reflectivity with respect to frequency.

\begin{figure}[htb]
\includegraphics[width= \columnwidth]{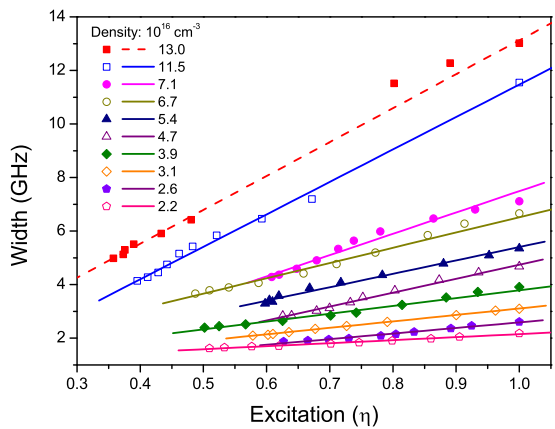}%
\caption{(Color online) The fitting value of the width is plotted as
a function of the excitation factor $\eta$. Different colors
represent the results at different atomic density from N=$1.3 \times
10^{17}$ cm$^{-3}$ to N=$2.2 \times 10^{16}$ cm$^{-3}$. The straight
lines are the linear fits.} \label{fig:alldensity}
\end{figure}

In order to obtain the width and excitation factor from the
experimental data, we use the expression of FM spectra to fit the
experiment data by leaving the self-broadened width $\Gamma$, the
excitation factor $\eta$ and the line shift $\Delta \Omega$ as
fitting parameters. The excitation factor $\eta$ is normalized to
unity for the case where no pump laser is applied. In our
experiment, Rb vapor contains natural abundance of $^{85}$Rb and
$^{87}$Rb which gives rise to four doppler-broadened absorption
lines in the absorption spectra. All of four components are taken
into account in the fitting of reflection spectra. Each component is
given a normalized oscillator strength. The dashed curves in Fig.
\ref{fig:narrow} are the examples of the fitting. For the case
without the pump laser, the fitted width is 13.0$\pm$0.3 GHz and
$\eta = 1.0$; for the case with the pump laser (laser power P=180
mW), the fitted width is 4.98$\pm$0.05 GHz and $\eta=0.36$. At the
same atomic density (N=$1.3 \times 10^{17}$ cm$^{-3}$), the FM
spectrum is recorded when we apply the pump laser with different
powers. The fitting of these FM spectra gives the widths
corresponding to the different excitation factors. The fitting
results are shown as the red squares in Fig. \ref{fig:alldensity},
where the width is plotted as a function of the excitation factor
$\eta$. The dashed line in Fig. \ref{fig:alldensity} is a linear fit
($y=a+b x$, where the dependent variable $x$ represents the
excitation $\eta$ and $a$ and $b$ are fitting parameters) of the
excitation dependence of the width at atomic density N=$1.3 \times
10^{17}$ cm$^{-3}$, and the slope (fitting parameter $b$) is 12.7
GHz.

\begin{figure}[htb]
\includegraphics[width= \columnwidth]{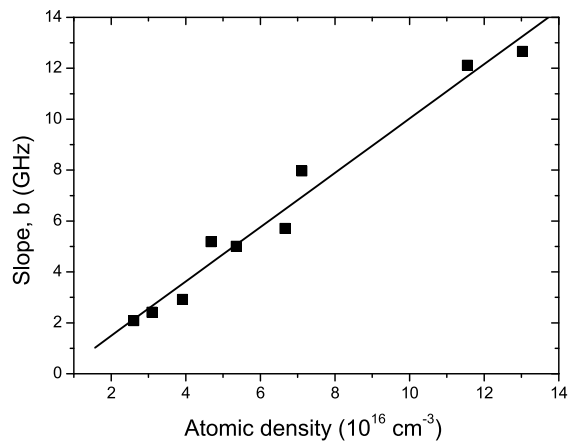}%
\caption{The slope of the width dependence on the atomic density is
plotted as a function of the atomic density. Squares are the results
of fitting from Fig. \ref{fig:alldensity}. The solid straight line
is a linear fit.} \label{fig:slope}
\end{figure}

\begin{figure}[tb]
\includegraphics[width= \columnwidth]{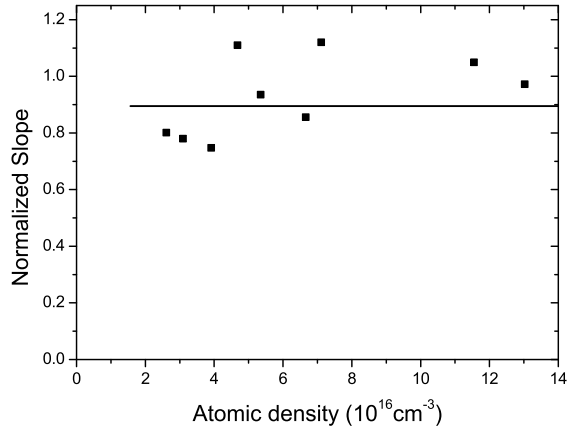}%
\caption{Normalized slope is plotted as a function of the atomic
density. Squares are the data extracted from experimental results.
The solid line is the linear fit with a zero slope. The fit value of
the normalized slope is 0.90$\pm$0.05.} \label{fig:nslope}
\end{figure}

We determine the widths and excitation factors by measuring and
fitting the FM reflection spectra at different atomic densities from
N=$1.3 \times 10^{17}$ cm$^{-3}$ to N=$2.2 \times 10^{16}$
cm$^{-3}$. The measured density dependence of the self-broadened
rubidium D$_2$ linewidth is the same as in Refs. \cite{Kondo2006}.
The excitation dependence of the width for different atomic
densities is shown in Fig. \ref{fig:alldensity} with different
colors. The corresponding linear fit gives the slope (width/$\eta$)
for each density. In Fig. \ref{fig:slope}, the slope is plotted as a
function of the atomic density. The solid line is a linear fit. If
the slope for each density is normalized by taking the ratio of the
slope to the width without the pump laser at each density, the
normalized slopes are close to unity. As shown in Fig.
\ref{fig:nslope}, the value of the normalized slope is 0.90$\pm
$0.05. According to our simple model for the fit, the normalized
slope is density independent. It indicates that the self-broadening
at high atomic density is a combination of collision and static
spectral broadening (inhomogeneous profile).

\
\
\
\

\section{Conclusion}

\ We have observed that the excitation dependence of self-broadening
is the same in the range of atomic density from $2.2\times 10^{16}$
to $1.3\times 10^{17}$ cm$^{-3}$, where dipole-dipole interactions
are a dominant source of spectral broadening. Our results support
the predictions of the theoretical model developed in Ref.
\cite{Mukamel1994}. The obtained results can be useful for
understanding of excitation processes in a condensed media such as
solutions, glasses, polymers, proteins and molecular crystals. In
atomic gases it will be interesting to study a possible transition
from many body interactions (disordered excitons) \cite{Mukamel1994,
Cundiff2008, VAS1996} to the resonance two-body interaction (impact
collisions) at lower atomic densities \cite{Lewis, Vdovin}. Probably
the studies will require applications of non-linear optical methods
in frequency and time domain as the complimentary spectroscopic
techniques. We shall note that additional information about the
dipole-dipole interaction in a dense gas may be obtained by using a
nano-cell \cite{Picher2008}.
\section{Acknowledgement}

\ We thank S.T. Cundiff, S. Mukamel, D. Sarkisyan and R.W. Welch for
useful discussions. Also, we wish to acknowledge D. Sarkisyan and T.
Varzhapetyan for the high-temperature cell. This work is supported
by the Welch Foundation (Grant A-1261) and the NSF grant EEC-0540832
(MIRTHE ERC).

% Create the reference section using BibTeX:


\begin{thebibliography}{99}
\bibitem{Lewis} E. L. Lewis, Phys. Rep. \textbf{58}, 1 (1980).

\bibitem{Vdovin} Yu. Vdovin and N.A. Dobrodeev, Sov. Phys. JETP. \textbf{18}, 544
(1969).

\bibitem{movre1980} M. Movre, G. Pichler, J. Phys. B \textbf{13}, 697
(1980); V. Horvatic, M. Movre, R. Beuc and C. Vadla, ibid
\textbf{26}, 3679 (1993).


\bibitem{Mukamel1994} J.A. Leegwater and S. Mukamel, Phys. Rev. A \textbf{49}, 146
(1994).

\bibitem{Mukamel1995} S. Mukamel, \emph{Principles of nonlinear optical
spectroscopy}, (Oxford University Press, New York, 1995).

\bibitem{Mukamel1991} Y.J. Yan and S. Mukamel, J. Chem. Phys. \textbf{94},
179 (1991).


\bibitem{Cundiff2008} V. O. Lorenz and S. T. Cundiff, Phys. Rev. Lett. \textbf{95},
163601 (2005); V. O. Lorenz, S. Mukamel, W. Zhuang, and S. T.
Cundiff, Phys. Rev. Lett. \textbf{100}, 013603 (2008).

\bibitem{VAS1996} V. A. Sautenkov, H. van Kampen, E. R. Eliel, and J. P.
Woerdman, Phys. Rev. Lett. \textbf{77}, 3327 (1996).

\bibitem{HebinLi2008} H. Li, T.S. Varzhapetyan, V. A.
Sautenkov, Y.V. Rostovtsev, H. Chen, D. Sarkisyan, M.O. Scully,
Appl. Phys. B \textbf{91}, 229 (2008).

\bibitem{VAS1999} H. van Kampen, V. A. Sautenkov, C. J. C. Smeets, E. R.
Eliel and J.P. Woerdman, Phys. Rev. A \textbf{59}, 271 (1999).

\bibitem{VAS2008} V. A. Sautenkov, Y. V. Rostovtsev, E. R. Eliel,
Phys. Rev. A \textbf{78}, 013802 (2008).

%LOCAL FIELD CORRECTION IN EXPRESSIONS FOR R
\bibitem{Friedberg1973} R. Friedberg, S. R. Hartmann and J. T. Manassah,
Phys. Rep. \textbf{7}, 101 (1973); Phys. Rev. A \textbf{42}, 5573
(1990).

\bibitem{Manassah1983} J. T. Manassah, Phys. Rep. \textbf{101}, 359 (1983).

\bibitem{Maki1992} J. J. Maki, M. S. Malcuit, J. E. Sipe, and R. W. Boyd, Phys. Rev. Lett.
\textbf{67}, 972 (1991); J. J. Maki, W. V. Davis, R. W. Boyd, and J.
E. Sipe, Phys. Rev. A \textbf{46}, 7155 (1992).

\bibitem{Kondo2006} K. Niemax, M. Movre, and G. Pichler, J. Phys. B
\textbf{12}, 3503 (1979); R. Kondo, S. Tojo, T. Fujimoto, and M.
Hasuo, Phys. Rev. A \textbf{73}, 062504 (2006).

\bibitem{Picher2008} T. Varzhapetyan1, A. Nersisyan1, V. Babushkin1, D. Sarkisyan1, S.
Vdovi\'{c}
and G. Pichler, J. Phys. B \textbf{41}, 185004 (2008).


\end{thebibliography}
\end{document}